\documentclass[10pt]{article}
\usepackage{times}
\usepackage{natbib}
\title{A ``Lorentz-Poincar\'e"--Type Interpretation\\ of  Relativistic Gravitation}
\author{Jan (B.) Broekaert\footnote{CLEA-FUND, Vrije Universiteit Brussel,  Belgium; email:jbroekae@vub.ac.be}}

\begin{document}

\maketitle

\begin{abstract}
The nature of  \emph{time}, \emph{space} and  \emph{reality} are to large extent dependent on our interpretation of Special (SRT) and General Relativity Theory (GRT).  In SRT essentially two distinct  interpretations  exist;  the  ``geometrical" interpretation by Einstein based on the \emph{Principle of Relativity} and the \emph{Invariance of the velocity of light}  and, the  ``physical"  Lorentz-Poincar\'e  interpretation with underpinning by \emph{rod contractions}, \emph{clock slowing} and \emph{light synchronization}, see e.g. \citet{Bohm1965, Bell1987}.   It can be questioned whether the ``Lorentz-Poincar\'e"-interpretation of SRT  can be continued into GRT.  We have shown that  till first Post-Newtonian order this is indeed possible \citep{Broekaert2004}.  This requires the introduction of gravitationally modified Lorentz transformations, with  an intrinsical  spatially-variable speed of light $c(r)$, a  scalar scaling field  $\Phi$ and induced velocity field  $\mathbf{w}$. Still the invariance of the locally observed velocity of light is maintained \citep{Broekaert2005}. The Hamiltonian description of particles  and  photons recovers the 1-PN   approximation  of GRT.  At present we show the model does obey the Weak Equivalence Principle from a fixed perspective, and that the implied  acceleration transformations are equivalent with those of GRT. 
\end{abstract}

\section{Introduction}
We believe that a Lorentz-Poincar\'e (L-P) interpretation of relativistic gravitation has the merit of retaining a classical ontology of gravitational fields, and Hamiltonian mechanics in a flat-metric space.  However, proper to  an L-P interpretation of gravity, the physically observed measurements of space and time still result in a curved space-time. Basically this is the idea of  Poincar\'e's geometric conventionalism \citep{Poincare1902}: it is formally indistinguishable to have free geodesic motion in curved space-time, or to have and adjusted gravitational dynamics ---which also affects electromagnetism in  rods and clocks--- in a flat space and cosmic time.
The  present scalar--vector model is accordingly based on isotropic scaling, i.e. contraction and dilation, of physical quantities depending on  position in the gravitation  field and directional scaling due to velocity relative to gravitational sources. \\
Two levels of description must be discerned in a L-P type model (see also \citet{Cavalleri1980,Thirring1961, Dicke1957, Wilson1921}): gravitationally affected observations (or scaled) versus gravitationally unaffected observation (or unscaled).   Our model implements  the gravitational effects on rods and clocks through  appropriate gravitationally modified  Lorentz Transformations (GMLT's) for space and time. These relate  both, affected and unaffected observers,  but also by combination of the previous; distinctly affected observers.  Moreover the GMLT's for energy and momentum provide the hamiltonian expressions  which cover adequately the  gravitational phenomenology of GRT \citep{Broekaert2005}.
While the latter has been verified till 1-PN, some aspects of the model remain to be verified.\\
A critical requirement of a viable formulation of gravitation is the sufficient fulfillment  of the Equivalence Principle (EP) even more the Weak Equivalence Principle (WEP) \citep{Will1995} (see however also \citet{Damour2001}). The WEP  purports the local indistinguishability of acceleration and gravitation, or the equivalence of inertial and gravitational mass. The EP on the other hand requires al physical laws in local  free-falling  frames to be equivalent and, equivalent with  unaccelerated frames  in a homogeneous (zero) gravity field. Similarly  the WEP can be stated as the principle of universality of free-fall; and the free-falling observer locally observes  gravitation to be eliminated.  \\
The WEP should be apparent to a fixed affected observer as well. In that case the acceleration of the particle should invariantly be independent of its rest mass and energy. This configuration can be easily covered by the  acceleration transformations derived from the space-time GMLT's. We will detail in Section 3, that in this configuration  there is  a residual  relative acceleration which is  dependent on the   velocity of the particle. It will be shown that  the GRT expression of this acceleration is the same as in our L-P type model (Section 4). With the independence of the acceleration on the  mass of the free-falling system (Section 2),  this  proves the validity of the WEP for fixed observers in our Lorentz-Poincar\'e model in the same manner as in General Relativity Theory.

\section{The Lorentz-Poincar\'e type interpretation}
A physical observer at coordinate position $\mathbf{r}$ will locally measure affected quantities $(d\mathbf{x}', dt')$. The explicit structure of the Lorentz-transformation adapted to gravitation straightforwardly reveals the effects of gravitational rod contraction, clock slowing and synchronization at the given location,
\begin{eqnarray}
d{\bf x}'  &=&  \left(( {d{{\bf x}_o}_\parallel} -{{\bf u}_o}  d t_o)  {\gamma  \left( u_o \right)}     +
  {d{{\bf x}_o}_\perp} \right)  \Phi   \left( {\bf r} \right)^{-1}   \label{SotoSoacGMLTa} \\
d t' &=& \left( d t_o-   {{\bf u}_o}. d{{\bf x}_o}  {c_o ({\bf r})}^{-2}  \right)  {\gamma  \left( u_o \right)}  {\Phi   \left( {\bf r} \right)}   \label{SotoSoacGMLTb} 
\end{eqnarray}
which relates them to the unaffected space and time intervals of coordinate space. These are to be obtained taking into account the induced velocity field ---due to source movement--- in a Galilean relation in coordinate space:
\begin{eqnarray}
d\mathbf{x} =  d\mathbf{x}_o + \mathbf{w} dt_o  &, &    dt = dt_o 
\end{eqnarray}
The frame velocity  $\mathbf{u}$ of the affected observer relative to coordinate space, is then related to $\mathbf{u}_o$ by $\mathbf{u}_o = \mathbf{u} - \mathbf{w}$, while the velocity of light is given by $ c_o =  \vert \mathbf{c}_w - \mathbf{w} \vert$. The specific form of the relation (\ref{SotoSoacGMLTa}, \ref{SotoSoacGMLTb}) is based  on $\mathbf{u}_o$ and  $\gamma(\mathbf{u}_o)$ appearing as an effective dynamical velocity and relativistic \emph{kinematical} factor in the associated Hamiltonian mechanics and, concomitantly, the Lorentz-transformation-\emph{form} which constrains the velocities to comply to the invariance of the locally observed velocity of light.  \\
The associated velocity transformation is  straightforward:
\begin{eqnarray}
 {\mathbf v}' & = & \frac{ {{\bf v}_o}_\parallel - {\bf u}_o + \gamma^{-1}  {{\bf v}_o}_\perp}{1 -{\bf u}_o.{\bf v}_o {c_o}^{-2}}  \frac{1}{\Phi^2}  \label{velocityrelation}
\end{eqnarray}\\
The gravitational scaling and induced velocity fields $\{\Phi, \mathbf{w}\}$ are given by the equations:
\begin{eqnarray} 
  \Delta  \Phi  & = & \frac{4 \pi G' }{{c '}^2} \rho({\bf r} ) \Phi  +  \frac{\left(\nabla \Phi \right)^2}{\Phi}     \label{statgeneralPhi} \\
\Delta {\mathbf  w} &=&  -   \frac{16 \pi  G'}{{c'}^2}  \rho {\mathbf v}_\rho ({\mathbf x}, t) \label{wequation}
\end{eqnarray}
in no-retardation approximation. For example in the  static source configuration we find the well known solution:
\begin{eqnarray}
\Phi  \ \equiv \ \exp (\varphi) &,&  \ \varphi \ = \  { - \frac{ G'   }{ {c '}^2}  \int_S \frac{\rho({\bf r}^*)  }{\vert {\bf r} - {\bf r}^* \vert} d^3 r^*}  
\end{eqnarray}
The required hamiltonians are derived from associated energy-momentum GMLT's which expose gravitational affecting  different from the space-time GMLT's, and consistent with Newtonian fitting \citep{Broekaert2004}:
\begin{eqnarray}
H &=& m c^2 + \mathbf{p}. \mathbf{w}, \ \ \ m = m'_o \gamma (p) \Phi^{-3}
\end{eqnarray}
We have shown in previous work that the L-P model gives the correct 1-PN equations of motion for particles (and photons) in harmonic coordinate space: 
\begin{eqnarray}
\mathbf{a} &\approx& -{c'}^2 \nabla(\varphi + 2 \varphi^2)  -{v}^2 \nabla \varphi+ 4 \mathbf{v} \mathbf{v}.\nabla \varphi   - \mathbf{v} \times (\nabla \times \mathbf{w} ) + 3 \mathbf{v}\partial_t \varphi + \partial_t \mathbf{w}  \label{avec}
\end{eqnarray}
 In order to implement Poincar\'e's Principle of Relativity, and obtain the calibration of the $\mathbf{w}$-equation,  an acceleration transformation has been obtained;  reproduced here as Eq. (\ref{avectoavecac}).  Since the unaffected free-fall acceleration, Eq. (\ref{avec}),  is independent of rest-mass or energy,  this procedure ---to obtain the acceleration in affected perspective by transformation from the unaffected perspective--- trivially shows that the free-fall acceleration in the affected perspective as well does not  depend on the falling particle rest-mass or energy.  This independence shows then that the L-P model abides the  basic premiss of the Weak Equivalence Principle.   However it remains  to be verified whether whether this transformation correctly exposes  the velocity-dependent terms  in the  free-fall acceleration in the affected perspective of a fixed observer.

\section{Acceleration transformations in the Lorentz-Poincar\'e type model}
The acceleration transformation in the L-P model is obtained by taking the standard time derivative of the velocity transformation (\ref{velocityrelation}).  In the  case of a general kinematic source, $\dot \mathbf{w} \neq0$ and $\dot \Phi \neq 0$, the  acceleration observed by the affected observer is given by:
\begin{eqnarray}
{\mathbf{a}'} &=&   \left(1-  \mathbf{u}_o. \mathbf{v}_o c_o^{-2}\right)^{-2} \gamma_o^{-1} \Phi^{-3} 
 \left\{    {\mathbf{a}_o } -  \mathbf{o}_o  + (\gamma_o^{-1}  -1)   {\mathbf{a}_o }_\perp  +  {\mathbf{v}' }  {\Phi}^2  \mathbf{u}_o.{\mathbf{a}_o}    {c }_o^{-2 }   
\right.   \nonumber \\ 
& &   - \left( {\gamma }_o^{-1} -1 \right) u_o^{-2}
\left(     \mathbf{o}_o    \mathbf{v}_o .\mathbf{u}_o     + \mathbf{u}_o  \mathbf{v}_o.\mathbf{o}_o   - 2   \mathbf{o}_o . \mathbf{u}_o   {\mathbf{v}_o}_\parallel \right)   \nonumber \\ 
& & -  {\mathbf{v}_o }_\perp  {\gamma_o }    \mathbf{o}_o . \mathbf{u}_o  {c_o}^{-2}     + {\mathbf{v}' }  {\Phi}^2  \mathbf{o}_o.\mathbf{v}_o {c_o}^{-2 }   \nonumber \\ 
&& \left.- 2{\mathbf{v}' } {\Phi}^2 \left( 1 +   \mathbf{u}_o .{\mathbf{v}_o } {c_o} ^{-2}    \right)  
\dot{\varphi}_o        + 2  {\mathbf{v}_o }_\perp  {\gamma_o }    u_o^2{c}_o ^{-2 } \dot{\varphi}_o    \right\}  \label{avectoavecac}
\end{eqnarray}
with $\mathbf{o}=  \dot\mathbf{u}$ the observer-frame acceleration, $\mathbf{a}=  \dot\mathbf{v}$ the test-particle acceleration, both in terms of coordinate space, and  $\mathbf{a}'=  \dot\mathbf{v}'$ the test-particle acceleration in the perspective of the affected observer.  Effectively the nature of the affected observer  is completely defined by its frame particulars $\{\mathbf{u}, \dot\mathbf{u}\}$. The acceleration transformation can therefor be adapted to the affected observer being fixed ($\mathbf{u} = 0$, $\dot\mathbf{u}=0$) or  dragged  ($\mathbf{u} \neq 0$,$\dot\mathbf{u}=0$).  We compare the first   case  for the L-P model and GRT, while  other  cases, and the LIF-case in particular, will need to be elaborated in future work (see however \citet{Broekaert2005} for a specific dragged configuration).
\subsection{Acceleration relative to a fixed observer in the L-P model}
The  relative acceleration with respect to a fixed observer, i.e. \emph{frame} acceleration and velocity $\mathbf{o}  =  0$ and $\mathbf{u}  =  0$,  at 1-PN  is obtained  from Eq. (\ref{avectoavecac}) by approximation:
\begin{eqnarray}
 {\mathbf{a}'} &=&   \Phi^{-3} \left\{   {\mathbf{a}}  - 2{\mathbf{v}' } {\Phi}^3  \dot{\varphi}'     \right\} \label{LPcorrections}
\end{eqnarray}
where we have made use of  the  contravariant space-time GMLT, {$S'$} to {$S $} (the unaffected observer) for gradient operators:
\begin{eqnarray}
\nabla \ = \ \nabla_o & = &   \Phi^{-1} \left( \gamma (     \nabla' _\parallel + {\mathbf{ u}'}  {c '}^{-2}    \partial_{t'})  +  \nabla'_\perp \right)    \label{gradientGMLT}  \\ 
 \partial_{t} + \mathbf{w}.\nabla\   = \ \partial_{t_o} & = & {\gamma } {\Phi } \left( \partial_{t'} +  {{\bf u}'} .\nabla'     \right)  \label{partialderivtGMLT}
\end{eqnarray}
Rendering all expressions of Eq. (\ref{avec}) explicitly  in terms of S', gives for $\mathbf{a}'$:
\begin{eqnarray}
{\mathbf{a}'}  &=    - ({c'}^2  + {v'}^2)  \nabla' \varphi    + 2 \mathbf{v}' \mathbf{v}'.\nabla' \varphi     +  \mathbf{v}'\partial_t' \varphi   -   \mathbf{v}' \times (\nabla' \times \mathbf{w}' )  +   \partial_t'  \mathbf{w} '  & \label{avecac}
\end{eqnarray}
where $\nabla  = \Phi^{-1}      \nabla'  $ and $\partial_t =  \Phi \partial_{t'}$ has been used for the fixed observer. We must consider next the same configuration in the framework of GRT.

\section{GRT and relative acceleration}
We first observe that the expression of the 3-acceleration of particles or photons in the affected perspective, \emph{i.e.} in the curved space-time of the  observer, is not commonly used in GRT due to its explicit dependence on the chosen metric.  The description of particle-motion is  obtained by setting zero the second  covariant proper-time derivative  of LIF-coordinates; the (null) geodesic equation. Appropriate coordinate transformations then lead to acceleration expressions in the physical coordinates  of an observer.  \\
A number of specific acceleration \emph{transformation} laws were  described in GRT by  \citet{RindlerMishra1993,Mishra1994} and,  in  generic (and Fermi--) coordinates \citep{MTW1973, Bini1995,BunchaftCarneiro1998}): 
\begin{eqnarray}
\mathbf{a} &=& \mathbf{g} - \mathbf{g}. \mathbf{v} \mathbf{v} 
\end{eqnarray}
where $\mathbf{a}$ is the  local 3-proper-acceleration of a relativistic  particle  in a static gravitational field, relative to a stationary observer, and $\mathbf{g}$  is this same acceleration but with the ``physical" relative  velocity  $\mathbf{v} = 0$. \\
In order to have correspondence with the L-P model we compare to the expressions in
Schwarzschild coordinates  by \citet{McGruder1982}.  For the   affected, physically observed, radial acceleration  in GRT, McGruder gives:
\begin{eqnarray}
a_l  &=& g \left( v_R^2 - v_l^2 -1 \right) + O (r^{-3}) \label{RadAccSchwarz}
\end{eqnarray}
with $v_R$ the affected radial velocity and  $v_l$ the affected transversal velocity (here $g = {c'}^2 \kappa/r^2$). The comparison to the L-P model requires however that we work with its associated 1PN metric ;
\begin{eqnarray}
c'^2 {dt'_o}^2 - {d\mathbf{x}'_o}^2 & = &   \Phi^2  c'^2 {dt}^2 (1-w^2/c^4) -  \Phi^{-2}(d\mathbf{x} - \mathbf{w} dt)^2  \label{invariant}
\end{eqnarray}
then for  ${x'_o}^\mu = {x'_o}^\mu (x)$;
\begin{eqnarray}
g_{\mu \nu}  &=& \left( \begin{array}{cc}
\Phi^2 (1-{w'}^2{c'}^{-2})& \mathbf{w} \Phi^{-2}\\ 
\mathbf{w} \Phi^{-2} & - \Phi^{-2} \delta_{ij}
 \end{array} \right)   \label{GMLTmetric}
\end{eqnarray}
In the stationary source case ($\mathbf{w}=0$) this is the \emph{isotropic} Schwarzschild metric (see e.g. \citet{Ni1957}). In this case we find precise correspondence between L-P expression  (\ref{avecac}) and GRT expression (\ref{RadAccSchwarz}) for radial acceleration. 
In order to compare for the more general case we   calculate the full acceleration for a kinematical source, $\dot\mathbf{w}\neq 0$ and $ \dot \Phi \neq 0$, but still for a static observer.

\subsection{ Acceleration  relative to a fixed observer in GRT}
From the metric  tensor (\ref{GMLTmetric}) and its generic formula (see e.g.  \citet{Weinberg1972} Section 3.2):
\begin{eqnarray}
g_{\mu \nu}  &=&  \frac{\partial \xi^\alpha}{\partial x ^\mu} \frac{\partial \xi^\beta}{\partial x ^\nu} \eta_{\alpha\beta}
\end{eqnarray}
where $\xi^\alpha$ are some LIF-coordinates and  $x^\mu$ are taken in coordinate space,  we find till 1PN:
\begin{eqnarray}
b^\lambda_{\ \mu}  =    \frac{\partial \xi^{\lambda} (x) }{\partial x^\mu}  =  \left( \begin{array}{cc}
\Phi& 0\\ 
- \mathbf{w} &   \Phi^{-1} \delta_{ij}
 \end{array} \right)  \label{scalingrelation}
\end{eqnarray} 
which are defined up to a Lorentz boost $ \Lambda^\lambda_\beta (u')$ between LIF's.  However, in the fixed observer case  a Lorentz boost is not present and the observer's coordinates are obtained directly through:
\begin{eqnarray}
d{x'}_o^\lambda&=&   b^\lambda_{\ \mu}  dx^\mu \label{coordinatetransformation}
\end{eqnarray}  
Which, in the case of  Eq. (\ref{scalingrelation}) with $\mathbf{w}=0$, would be a simple scaling transformation. The comparison with the acceleration expression  of the  L-P model requires that we confine the observers to the same kinematic situation; fixed relative to the coordinate space ($\mathbf{u}$ = 0), therefor the coordinates  ${x'}_o$ must in the end be expressed as (see Eq. \ref{invariant}):
\begin{eqnarray}
d\mathbf{x}'_o &=&   d\mathbf{x}' -\mathbf{w}' dt'\\
dt'_o &=& dt'   \label{galilei}
\end{eqnarray}
according the Galilean transformation which is sufficient given the 1-PN target.\\
The geodesic equation expressed in the coordinate frame gives the particle's equation of motion in terms of derivatives to proper time:
\begin{eqnarray}
\frac{D^2 x^{\mu}}{d \tau^2} \  =  \ 0  &=&  \frac{d^2 x^{\mu}}{d \tau^2}  + \Gamma^\mu_{\ \nu \lambda} (x)\frac{d x^\lambda}{d \tau}  \frac{d x^{\nu}}{d \tau}  
\end{eqnarray}
or in terms of derivatives to coordinate time (\citet{Weinberg1972}, Eq. 9.1.2):
\begin{eqnarray}
\frac{d^2 x^{i}}{d t^2}   =  - \Gamma^i_{\ \nu \lambda}  \frac{d x^\lambda}{d t}  \frac{d x^{\nu}}{d t} + \Gamma^0_{\ \nu \lambda}  \frac{d x^\lambda}{d t}  \frac{d x^{\nu}}{d t}    \frac{d x^{i}}{d t}   &&\label{GRTdyn}
\end{eqnarray}
Till 1PN ---as we have mentioned before--- the resulting coordinate acceleration $\mathbf{a}$ is given by Eq. (\ref{avec}). In terms of the physical coordinates of the fixed observer, Eq. (\ref{coordinatetransformation}) allows to express the second derivative according:
\begin{eqnarray}
 \frac{d^2 \mathbf{x}}{d t^2} & = &  \Phi \frac{d }{d t'} \left( \Phi^2 (\mathbf{v}'_o + \mathbf{w})  \right) \nonumber \\
&=& \ \Phi^3\mathbf{a}' + 2 \phi^3 \dot \varphi' \mathbf{v}'  
\end{eqnarray}
where in the last step  the expressions have been cast in terms of $\{\mathbf{x}', dt'\}$
 of the fixed affected observer according Eq. (\ref{galilei}).  No supplementary terms are occurring at 1PN in the ``force"-terms of Eq. (\ref{GRTdyn}) when expressed in   $\{\mathbf{x}', dt'\}$-coordinates.  \\ We therefor obtain precisely the same corrections to the acceleration in GRT as when the acceleration is expressed in physical coordinates in the L-P model; Eq. (\ref{LPcorrections}). The free-fall acceleration relative to a fixed observer in GRT  is therefor, at 1-PN, precisely given by the acceleration (\ref{avecac}) of the  L-P model.

\section{The WEP in the Lorentz-Poincar\'e Model, conclusions}
We have shown ---under restricted conditions of observation--- the validity of the Weak Equivalence Principle in the L-P model and, detailed aspects of its formulation. \\
In particular we have shown that the free-fall acceleration  observed by a fixed observer in a general gravitation field, according the L-P model and in GRT, coincides  at first  Post-Newtonian order.  The WEP should of course be verified in more general conditions;   the kinematics of the  the observer should be generalized.  However, due to the fact that the L-P model uses an \emph{acceleration} transformation in order to obtain the free-fall acceleration in physical perspective, the independence of the rest mass is trivial, since the coordinate-space expression of the acceleration itself is independent of  the rest mass. The successful implementation of the WEP in the L-P model   therefor hinges on the specific aspects of an acceleration transformation of the type Eq. (\ref{avectoavecac}). It remains to be studied whether or not  adaptations  to it are required to cover LIF observations in concordance with the Weak Equivalence Principle.


\begin{thebibliography}{}

\bibitem[Bell(1987)]{Bell1987}
Bell J. S., \emph{Speakable and Unspeakable in Quantum Mechanics}, Cambridge University Press, 1987

\bibitem[Bini \emph{et al.}(1995)]{Bini1995}
Bini  D.,  Carini  P and Jantzen R.T., Relative observer kinematics in general relativity,
\emph{Classical and Quantum Gravity}, {\bf 12}, 2549-2563, 1995

\bibitem[Bohm(1965)]{Bohm1965}
 Bohm D., \emph{The Special Theory of Relativity},   W.A. Benjamin Inc,  1965 

\bibitem[Broekaert(2004)]{Broekaert2004} 
Broekaert J., A Spatially-VSL  gravity model with 1-PN limit of GRT, 2004, gr-qc/0405015 (submitted)

\bibitem[Broekaert(2005)]{Broekaert2005}
Broekaert J., A Modified-Lorentz-Transformation based Gravity Model Confirming Basic GRT-Experiments, \emph{Foundations of Physics}, {\bf 35}, 839-864, 2005

\bibitem[Bunchaft, Carneiro(1998)]{BunchaftCarneiro1998}
Bunchaft F., Carneiro S., The static spacetime relative acceleration for the general free fall and its possible experimental test, \emph{Classical and Quantum Gravity}, {\bf 15}, 1557,  1998

\bibitem[Cavalleri, Spinelli(1980)]{Cavalleri1980}
Cavalleri G.,  Spinelli G.,  Field-theoretic approach to gravity in flat space-time, \emph{La Rivista del Nuovo Cimento}, {\bf 3}, 8, 1980

\bibitem[Damour(2001)]{Damour2001}
Damour T., Questioning the Equivalence Principle,  2001, gr-qc/0109063

\bibitem[Dicke(1957)]{Dicke1957}  
Dicke  R.H., Gravitation without a {P}rinciple of {E}quivalence, \emph{Reviews of Modern Physics}, {\bf 29}, 363-376, 1957 

\bibitem[McGruder(1982)]{McGruder1982}
McGruder III, Ch. H.,  Gravitational Repulsion in the Scwarzschild field,  \emph{Physical Review D}, {\bf 25}, 3191-3194, 1982

\bibitem[Mishra(1994)]{Mishra1994}
Mishra L.,  The relativistic acceleration addition theorem, \emph{Classical and Quantum Gravity}, {\bf 11}, L97 -L102, 1994

\bibitem[Misner \emph{et al.}(1973)]{MTW1973}
Misner C.W, Thorne K.S.,  Wheeler J.A., \emph{Gravitation}, W.H. Freeman, San Francisco,1973

\bibitem[Ni(1957)]{Ni1957}  
Ni W.T.,  Theoretical Frameworks For Testing Relativistic Gravity. IV. A Compendium of Metric Theories of Gravity and their Post-Newtonian Limits, \emph{Astrophysical Journal}, {\bf 176}, 769-796, 1972 


\bibitem[Poincar\'e(1902)]{Poincare1902}
Poincar\'e H., \emph{La Science et l'Hypoth\'ese}, Edition Flammarion, Paris, 1902: 1968

\bibitem[Rindler, Mishra(1993)]{RindlerMishra1993}
Rindler W.,  Mishra L., The nonreciprocity of relative acceleration in relativity, \emph{Physics Letters A}, {\bf 173}, 105-108, 1993

\bibitem[Thirring(1961)]{Thirring1961} 
 Thirring T.E.,  An {A}lternative {A}pproach to the {T}heory of {G}ravitation, \emph{Annals of Physics}, {\bf 16}, 96-117, 1961


\bibitem[Weinberg(1972)]{Weinberg1972}
Weinberg S., \emph{Gravitation and {C}osmology. {P}rinciples and applications of the {G}eneral {T}heory of {R}elativity},  Wiley, London, 1972.


\bibitem[Will(1995)]{Will1995}
Will C., \emph{Theory and Experiment in Gravitational Physics}, Cambridge University Press,1995

\bibitem[Wilson(1921)]{Wilson1921}
Wilson  H.A.,  An Electromagnetic Theory of Gravitation, \emph{Physical Review}, {\bf 17}, 54-59, 1921

\end{thebibliography}
\end{document}